# Is Jobless Growth Valid in Türkiye? A Sectoral Analysis of the Relationship between Unemployment and Economic Growth

Emre AKUSTA*

**ABSTRACT**

This study analyzes the validity of jobless growth in Türkiye on sectoral basis. It analyzes the impacts of agriculture, industry, construction and services sectors on unemployment using annual data for the period 2000-2022. ARDL method is applied within the scope of the analysis. The findings are tested with FMOLS and CCR methods. The results show that growth in all sectors reduces the unemployment. A one-unit increase in the share of agriculture sector in GDP decreases the unemployment rate by 0.471 points, 0.680 points in the industrial sector, 0.899 points in the construction sector and 1.383 points in the services sector in the short-run. The long-run coefficients reveal that the impacts of sectoral growth on unemployment are stronger in the long-run than in the short-run. A one unit increase in the share of the agricultural sector in GDP decreases the unemployment rate by 2.380 points, 4.057 points in the industrial sector, 1.761 points in the construction sector and 3.664 points in the services sector in the long-run. These findings show that jobless growth is not valid in Türkiye in general. On the contrary, economic growth plays an important role in reducing unemployment.

# Türkiye'de İstihdam Yaratmayan Büyüme Geçerli mi? İşsizlik ve Ekonomik Büyüme İlişkisinin Sektörel Analizi

**ÖZ**

Bu çalışma, Türkiye'de istihdam yaratmayan büyümenin geçerliliğini sektörel düzeyde analiz etmektedir. 2000-2022 dönemine ait yıllık veriler kullanılarak tarım, sanayi, inşaat ve hizmetler sektörlerinin işsizlik oranı üzerindeki etkileri incelenmiştir. Analiz kapsamında ARDL yöntemi uygulanmıştır. Elde edilen bulgular FMOLS ve CCR yöntemleri ile sınanmıştır. Sonuçlar, tüm sektörlerde büyümenin işsizlik oranını azalttığını göstermektedir. Tarım sektörünün GSYH içindeki payındaki bir birim artış, kısa vadede işsizlik oranı 0.471 puan, sanayi sektöründe 0.680 puan, inşaat sektörünün 0.899 puan ve hizmetler sektöründe ise 1.383 puan düşürmektedir. Uzun dönem katsayıları, sektörel büyümenin işsizlik üzerindeki etkilerinin kısa döneme kıyasla daha güçlü olduğunu ortaya koymaktadır. Tarım sektörünün GSYH içindeki payındaki bir birim artış, işsizlik oranını uzun vadede 2.380 puan, sanayi sektöründe 4.057 puan, inşaat sektöründe 1.761 puan ve hizmet sektöründe 3.664 puan azaltmaktadır. Bulgular, Türkiye'de genel olarak istihdamsız büyümenin geçerli olmadığını göstermektedir. Aksine, ekonomik büyüme işsizliğin azaltılmasında önemli bir rol oynamaktadır.

**Anahtar Kelimeler:** İstihdam yaratmayan büyüme, Sektörel büyüme, ARDL, FMOLS, CCR

**JEL Sınıflandırması:** E24, J64, O40

---

* Assist. Prof. Dr., Kırklareli University, Faculty of Economics and Administrative Sciences, Department of Economics, emre.akusta@klu.edu.tr, ORCID:0000-0002-6147-5443

# 1. INTRODUCTION

Economic growth and employment are among the most fundamental macroeconomic indicators that determine the success of a country's economy. While growth refers to the increase in the total production volume of an economy, employment refers to the amount of labor force used in this production process. It is widely accepted in the economics literature that economic growth is expected to lead to an increase in employment. The expansion in production capacity should increase the general welfare of society by creating new job opportunities and reducing unemployment rates (Jawad & Naz, 2024). However, the impact of growth on employment varies from country to country and from sector to sector (Haider et al., 2023). The capacity of growth to create employment may differ depending on the nature of economic policies, technological developments and sectoral transformations.

One of the main objectives of growth is to increase the welfare of the society and to use the labor force efficiently. However, there may be situations where growth alone is not sufficient. Especially if the increase in production is not synchronized with the increase in employment, the quality of growth becomes questionable (Bogoviz et al., 2018). The United Nations Human Development Report of 1996 categorized different types of growth. In this report, jobless growth was categorized as one of the types of growth to be avoided (UNDP, 1996). The failure of economic growth to create jobs is seen as a serious problem, especially in developing economies. Jobless growth means that unemployment rates do not fall or employment growth remains limited despite positive growth rates in an economy (Haider et al., 2023). This phenomenon indicates that the relationship between economic growth and employment has weakened. Such a situation calls into question the effectiveness of economic policies. Globalization, technological transformations and capital-intensive production models are among the main reasons for this process.

One of the main causes of jobless growth is technological developments. Modern production techniques increase labor productivity and reduce labor demand. Especially automation and technological innovations minimize the need for labor in many sectors. The impacts of technological developments on employment have been discussed since the industrial revolution. Although such innovations were initially thought to increase employment, they are now seen to have an increasing impact on unemployment rates. Especially low-skilled labor is adversely affected by this transformation and imbalances in the labor market are deepening (Saratchand, 2019; Damioli et al., 2024). Another important reason is globalization. Increasing global competition forces firms to make their production processes more flexible. Therefore, firms reduce costs and aim to gain competitive advantage by reducing labor costs. This trend, especially in the industrial sector, leads to a shift of production to countries with low labor costs. This leads to high unemployment rates in countries (Davidson et al., 2020). However, global supply chains create new job opportunities in some sectors. However, the capacity of such growth models to create employment remains limited.

Another reason for jobless growth is changes in the structure of production. Especially in developing countries, the sectoral composition of growth is of great importance. The transition from agriculture to industry and services leads to changes in the employment structure. However, since the industrial sector is generally capital intensive, such growth models increase labor demand to a limited extent. Similarly, as the share of the services sector in the economy increases, the demand for labor increases directly. However, employment in these sectors is characterized by flexible working patterns and low wages. Therefore, employment growth in the services sector does not always fully reflect the positive impact of economic growth on employment (Ghosh, 2011; Oyelaran-Oyeyinka & Lal, 2016; Behera, 2019). Moreover, government policies and economic stability also play a decisive role in the failure of economic growth to create jobs. Factors such as tax policies, labor market regulations and investment incentives directly affect employers' propensity to create new jobs (Murat & Yılmaz Eser, 2013). In particular, high labor costs may lead employers to hire fewer employees. Such a situation may limit the capacity of growth to reduce unemployment. Moreover, economic crises and uncertainties may adversely affect firms' long-term employment decisions and reinforce the process of jobless growth.

The Türkiye's economy has achieved high growth rates, especially after the 2001 economic crisis. However, the impact of this growth on employment has varied periodically and sectorally (Herr & Sonat, 2014). Structural transformations in the industrial sector and the foreign-dependent production model weakened the impact of growth on employment. Shifts in employment occurred during the transition from agriculture to industry. However, the failure to create sufficient new job opportunities has kept unemployment rates high (Karapinar, 2007; Pehlivanoglu & Ince, 2020). This situation affects the structural factors that determine the relationship between growth and employment. Therefore, the extent to which Türkiye has adapted to job-creating growth policies should be investigated. In this regard, analyzing growth in terms of its structural components and sectoral dynamics is a critical requirement to assess the validity of jobless growth in the Turkish economy. In the literature, the relationship between economic growth and employment in Türkiye is frequently analyzed at the general economy level. However, this approach ignores sectoral dynamics. However, it is of great importance to consider sectoral differences in order to analyze employment dynamics more effectively. Therefore, this study re-examines the validity of jobless growth in Türkiye at the sectoral level. A sub-sectoral analysis will contribute to the development of policy recommendations for increasing employment. In particular, employment trends in agriculture, industry, services and construction sectors will be evaluated comparatively. Finally, recommendations for Türkiye's job-creating growth strategies will be presented.

This study contributes to the literature in at least four main aspects: (1) Unemployment growth in Türkiye is usually analyzed at the level of total economic growth. However, sectoral analyses are quite limited. This study aims to contribute to the literature in this area. (2) In this study, short-run and long-run coefficient estimates are obtained separately using the ARDL method. In this way, the relationship between sectoral growth and unemployment is analyzed in detail in both the short and long-run. (3) In order to increase the reliability of the results, the findings obtained with the ARDL method are tested with FMOLS and CCR methods. Therefore, robustness checks were performed. Comparisons with different econometric methods support the consistency and reliability of the findings of the study. (4) The findings of the study provide important implications at the sectoral level. Moreover, policy recommendations are developed to support sustainable employment.

## 2. LITERATURE REVIEW

The relationship between economic growth and employment has been discussed in the literature for many years. In his pioneering study, Okun (1962) showed that there is an inverse relationship between growth and unemployment in the US economy and established the framework later known as "Okun's Law". However, studies conducted in different periods and countries have shown that this relationship has changed over time and is not uniform. Especially since the end of the 20th century, while growth rates have been high in some countries, employment growth has remained limited, revealing "jobless growth". Jobless growth has been addressed with different dimensions in developed and developing economies. Global studies reveal that the growth-employment relationship differs across countries. For this reason, international and Türkiye-specific studies are discussed separately in the following section.

## 2.1. Related Work in the International Context

International studies investigating jobless growth have examined the relationship between employment and growth from various perspectives. Some studies show that the relationship between unemployment and growth is affected by geographical and sectoral dynamics. Lal et al. (2010) examined the long-run relationship between growth and unemployment in Bangladesh, Pakistan, India, Sri Lanka and China. The results show that Okun's law is generally valid. Hamia (2016) analyzed 17 developing countries in the Middle East and North Africa. In this study, contradictory findings were obtained regarding the validity of Okun's law. These differences are explained by structural changes in the labor market. Hanusch (2013) finds that Okun's law is stronger in non-agricultural sectors in East Asian countries. However, this relationship is reversed in agricultural employment. These results show that the structure of the labor market shapes the relationship between growth and employment.

In some countries, the weakening of the impact of growth on unemployment is associated with sectoral transformations. Tejani (2016) argues that the elasticity between output growth and employment in India has decreased over time and that this is due to productivity gains. Pieper (2003) finds support for Verdoorn's law[1] and shows that the relationship between output growth and employment has weakened in developing countries. Onaran (2008) shows that the increase in unemployment in Central and Eastern European countries is more pronounced in the manufacturing sector. He attributed this to the impacts of foreign direct investment and international trade. However, the economic integration process could not prevent job losses. Technological progress is recognized as one of the most important causes of unemployment growth. Li et al. (2017) argue that the fourth industrial revolution can create new employment opportunities by increasing productivity. However, unemployment may increase due to the automation of manual jobs. Similarly, Peters (2017) shows that capital-intensive production combined with labor-saving technologies may have negative impacts on unemployment.

The impact of growth on employment in Europe has been analyzed in the context of crises and structural transformations. Jablanovic (2017) found that the unemployment rate in the Eurozone showed a continuous upward trend between 1991 and 2015. Kallioras et al. (2016) analyzed the negative impacts of post-crisis economic shocks and structural distortions on employment in Greece. Correira and Alves (2017) find that there has been a significant decline in employment rates in Portugal since the early 2000s. This trend accelerated especially after the 2008 global financial crisis. Similarly, Blázquez-Fernández et al. (2018) argue that the relationship between growth and employment in European Union countries differs across age and gender groups. In particular, high youth unemployment rates suggest that economic growth is not equally reflected across all demographic groups. Klinger and Weber (2020), meanwhile, find that after the 2008 global crisis in Germany, the job-creating capacity of growth declined and the upward trend in unemployment strengthened.

The impacts of globalization and FDI on employment in developing countries is also a controversial issue. Rizvi and Nishat (2009) find that the employment generating capacity of FDI in Pakistan, India and China is quite low. Mitra (2011) argues that the impact of trade on job creation in India is negligible. Das and Ray (2019), in a study on South Asian countries, find that FDI has no long-run impact on employment. However, panel data analysis shows that this relationship is valid in some countries. Das and Ray (2020) argue that the long-run relationship between globalization and employment is weak in India and some countries in South Asia. This suggests that while globalization may contribute positively to the labor market in some countries, it may lead to jobless growth in others.

---

[1] Verdoorn's Law was introduced by Verdoorn (1949) and developed by Kaldor (1966). It generally states that economic growth increases labor productivity. Especially in the manufacturing industry, production growth increases productivity through factors such as economies of scale, learning effects and technological developments.

Generally, the capacity of growth to create employment varies depending on the economic structure of countries, sectoral dynamics and changes in global markets. Therefore, job-creating growth policies should be developed based on the specific economic structure of each country. Moreover, sector-based strategies should be developed. Especially in developing countries, education policies need to be more aligned with the labor market and the employment generating capacity of their sectors needs to be increased.

**2.2. Related Work in Türkiye**

Studies on jobless growth in Türkiye have examined how this relationship is shaped in different periods and sectors. Beyazit (2004) emphasized that for economic growth to be sustainable, productivity increases should also increase employment. However, productivity increases alone may not be sufficient. Some studies have argued that there is no statistically significant relationship between growth and unemployment in Türkiye. Takim (2010) and, Korkmaz and Yılgör (2010) investigated the causality between growth and unemployment. In the post-2001 period, no causality relationship was found. These studies indicate that the capacity of growth to create employment in the Turkish economy may be limited. There are also some studies that reveal the existence of a negative relationship between growth and unemployment. For instance, Barisik et al. (2010) find an inverse relationship between growth and unemployment in Türkiye. This finding was later supported by Ozdemir and Yildirim (2013). However, most of these studies show that the impact of growth on reducing unemployment is stronger in periods of crisis and contraction, while this impact weakens in periods of expansion. Ceylan and Sahin (2010) find that the growth-unemployment relationship is asymmetric. In the study, it is stated that unemployment increases rapidly during contractionary periods but does not decrease at the same rate during expansionary periods. The main reasons for jobless growth include productivity gains, capital-intensive production structures and labor market rigidities. Altuntepe and Guner (2013) conducted a sectoral analysis for the 1988-2011 period and showed that the impact of growth on employment varies depending on sectoral differences. In particular, it is found that the employment creation potential of the services sector is higher than that of industry and agriculture.

In general, jobless growth occurs for different reasons in different countries. In advanced economies, factors such as global crises, technological transformations and capital-intensive production structures are cited as the main causes of jobless growth. In developing countries, on contrary, structural weaknesses in the labor market, productivity gains and the limited impact of factors such as foreign direct investments on job creation come to the fore. However, the level of development of a country is not the only factor determining the impact of growth on employment. The sectoral components of growth and the sectors in which labor demand is concentrated are also decisive. The capacity of growth to create employment is not at the same level in all sectors. The impacts of sub-sectors such as agriculture, industry and services on labor demand differ significantly. Even if growth accelerates in capital-intensive sectors, this increase may not be reflected in employment to the same extent. In labor-intensive sectors, however, an increase in production generally generates more employment. Therefore, in order to analyze the growth-employment relationship more effectively, it is necessary to make an assessment on a sectoral basis. Analyses conducted at the general economy level may be insufficient to explain the impact of growth on employment. In this scope, a detailed analysis of the components of growth is a critical step to test the validity of growth without employment in the Turkish economy. Detailed analyses by sub-sectors will contribute significantly to the development of job-creating growth policies.

In the literature, studies analyzing unemployment growth at the sectoral level in Türkiye are quite limited. For example, Tuncer and Altiok (2012) find that employment growth remains limited despite the growth in the manufacturing industry. Aksoy (2013), meanwhile, compares the relationship between growth and employment in the industrial and service sectors. The findings show that the employment creation impact of industrial growth is weaker. To the best of our knowledge, the only study that deals with this issue in detail is Abdioglu and Albayrak (2015). In this study, employment elasticity coefficients for the period 1988-2015 in Türkiye are estimated both at the general economy level and at the sub-sector level. The findings show that the highest employment elasticity is realized in the construction sector. While the impact of economic growth on employment is more pronounced in the construction sector, it is more limited in other sectors. Moreover, positive and negative output gaps do not make a significant difference on employment for most sectors.

This study aims to contribute to the limited literature by analyzing unemployment growth in Türkiye on a sectoral basis. In order to fill the existing gap in the literature, it provides up-to-date evidence using the most recent dataset. Moreover, it estimates both short-run and long-run coefficients using the ARDL method. Thus, unlike the literature, it reveals how the dynamics between employment and growth differ over time. Moreover, robustness checks are performed using FMOLS and CCR methods. Thus, the reliability of the results obtained is enhanced. The use of different econometric methods reinforces the accuracy of the model and confirms the robustness of the results. Finally, in addition to providing a theoretical analysis, our study also develops concrete recommendations. Thus, it is aimed to contribute to the development of strategies that will increase the employment generating capacity of growth.

## 3. DATA AND METHODOLOGY

### 3.1. Model Specification and Data

The empirical investigation of this study analyzes the validity of jobless growth by sectors[2] in Türkiye. For this purpose, the impacts of agriculture, industry, construction and services sectors on the unemployment rate are analyzed using annual data for the period 2000-2022. Most of the literature on jobless growth has been prepared using annual data. Therefore, annual data has been used in this study to ensure comparability with the literature and to increase the interpretability of the coefficients obtained. In addition, the inflation variable is also included in the model to provide a more comprehensive assessment of labor market dynamics. The period of analysis is chosen considering the availability of the data set. The empirical model of the study is shown in Equation 1.

$$UNP_t = \beta_0 + \beta_1 AGR_t + \beta_2 IND_t + \beta_3 CON_t + \beta_4 SER_t + \beta_5 INF_t + \epsilon_t \qquad (1)$$

In equation (1), $UNP$ is the unemployment rate, $AGR$ is the share of agriculture in GDP, $IND$ is the share of industry in GDP, $CON$ is the share of construction in GDP, $SER$ is the share of services in GDP and $INF$ is inflation measured by the consumer price index. Descriptive statistics of the variables are presented in Table 1.

---

[2] This study uses the classification provided by the Republic of Türkiye – Ministry of Treasury and Finance (2024).

**Table 1:** Descriptive Statistics

| Variables | Symbol | Description | Obs. | Mean. | Std.S. | Min. | Max. | Source |
|---|---|---|---|---|---|---|---|---|
| Agriculture Sector | AGR | % of GDP | 23 | 7.67 | 1.40 | 5.50 | 10.20 | TurkStat |
| Industry Sector | IND | % of GDP | 23 | 20.71 | 2.10 | 18.40 | 27.10 | TurkStat |
| Construction Sector | CON | % of GDP | 23 | 6.29 | 1.32 | 4.50 | 8.50 | TurkStat |
| Services Sector | SER | % of GDP | 23 | 53.93 | 1.26 | 51.20 | 57.20 | TurkStat |
| Unemployment | UNP | % of total labor force | 23 | 10.68 | 1.59 | 6.50 | 14.03 | World Bank |
| Inflation | INF | Consumer prices (%) | 23 | 18.71 | 18.30 | 6.25 | 72.31 | World Bank |

Note: (1) Obs., Mean, Std.S., Min, and Max denote observations, mean, standard deviation, minimum, and maximum, respectively. (2) TURKSTAT and World Bank represent the Turkish Statistical Institute – Gross Domestic Product Statistics and the World Bank – World Development Indicators (WDI), respectively.

Table 1 presents descriptive statistics of the variables used in the study. The services sector (SER) constitutes the largest part of GDP with an average share of 53.93%. The lowest value is 51.20% and the highest value is 57.20%. The low standard deviation indicates that the services sector has been relatively stable over the years. This indicates that the Türkiye's economy relies heavily on the services sector. The industrial sector (IND) has an average share of 20.71%. The lowest was recorded at 18.40% and the highest at 27.10%. The relatively high standard deviation indicates that fluctuations in the industrial sector are significant. The agricultural sector (AGR) has an average share of 7.67% of GDP. Its share varies between 5.50% and 10.20%. The agricultural sector is still important in terms of employment, but plays a more limited role in economic growth.

The construction sector (CON) accounts for 6.29% of GDP on average. The share of the construction sector in GDP varies between 4.50% and 8.50%. The low standard deviation indicates that fluctuations in the sector are relatively limited. Although the construction sector is the engine of growth periodically, it follows a more balanced course in the long-run. The unemployment rate (UNP) was calculated as 10.68% on average during the period analyzed. It also varied between 6.50% and 14.03%. The high standard deviation indicates that the unemployment rate fluctuates depending on economic conditions. The inflation rate (INF) was 18.71% on average. It reached a low of 6.25% and a high of 72.31%. The high standard deviation indicates that inflation experiences large fluctuations. In conclusion, the services sector plays a dominant role in the economy, while the industrial sector is also an important driver of growth. Agriculture and industry have smaller shares. Fluctuations in unemployment and inflation rates are indicators that need to be carefully monitored to ensure economic stability.

### 3.2. Unit Root Test

In time series analysis, determining whether the series are stationary or not is one of the fundamental steps of the analysis process. Stationarity means that the statistical properties of a series (such as mean, variance) do not change over time. Working with non-stationary series may lead to false results (spurious regression problem) (Granger & Newbold, 1974; Johansen & Juselius, 1990). Therefore, unit root tests are used to test the stationarity of the series. In this study, ADF (Augmented Dickey-Fuller) test and Phillips-Perron (PP) test are used.

The ADF test developed by Dickey and Fuller (1979) is one of the most widely used unit root tests. The ADF test is based on the regression equation in Equation 2:

$$\Delta Y_t = \alpha + \beta_t + \gamma Y_{t-1} + \sum_{i=1}^{p} \emptyset_i \Delta Y_{t-i} + \varepsilon_t \tag{2}$$

In Equation 2, $\Delta Y_t$ is the first difference of the series, $\alpha$ is the constant term, $\beta_t$ is the trend component, $\gamma$ is the unit root parameter, $\emptyset_i$ is the coefficients of the lagged differences and $\varepsilon_t$ is the error term. The ADF test tests the null hypothesis that $\gamma = 0$ (the existence of a unit root). If $\gamma$ is statistically significant and negative, the series is stationary.

The PP test, developed by Phillips and Perron (1988), tests for the presence of a unit root similar to the ADF test. However, the PP test corrects for autocorrelation and variance in the series using a nonparametric approach. The PP test is implemented through Equation 3 as follows:

$$\Delta Y_t = \alpha + \beta_t + \gamma Y_{t-1} + \varepsilon_t \tag{3}$$

In Equation 3, $\Delta Y_t$ is the first difference of the $Y_t$ series, $\alpha$ is the constant term, $\beta_t$ is the deterministic trend component, $\gamma$ is the unit root parameter, and $\varepsilon_t$ is the error term. Unlike the ADF test, the PP test does not add lagged differences. Instead, it uses Newey-West standard errors to correct for autocorrelation and heteroskedasticity in the error term. Both tests are used to determine whether the series contains a unit root and the results indicate the stationarity of the series.

### 3.3. Cointegration Test

The cointegration method is used to test whether there is a long-run relationship between variables. In this study, the ARDL (Autoregressive Distributed Lag) bounds test developed by Pesaran et al. (2001) is used to analyze the cointegration relationship. The ARDL test can be applied for both I(0) and I(I) level series. This makes it more flexible than other cointegration tests. The ARDL bounds test is based on the basic model in Equation 4:

$$\Delta Y_t = \alpha + \sum_{i=1}^{p} \beta_i \Delta Y_{t-i} + \sum_{i=0}^{q} \gamma_i \Delta X_{t-i} + \delta_1 Y_{t-1} + \delta_2 X_{t-1} + \varepsilon_t \tag{4}$$

In Equation 4, $Y_t$ is the dependent variable, $X_t$ is the independent variable, $\Delta$ is the difference operator, $p$ and $q$ are lag lengths, $\alpha$ is the constant term, $\beta_i$ and $\gamma_i$ are short-run coefficients, $\delta_1$ and $\delta_2$ are long-run coefficients, and $\varepsilon_t$ is the error term. The bounds test tests the hypothesis that $\delta_1 = \delta_2 = 0$. If the F-statistic exceeds the critical values, it is concluded that there is a long-run relationship between the variables.

The ARDL bounds test provides reliable results especially in small samples. Moreover, it allows modeling both short-run and long-run dynamics simultaneously. Therefore, it is widely used in time series analysis.

**3.4. Elasticities Estimator**

Once cointegration is detected, the ARDL model is used to estimate the long-run and short-run coefficients. Long-run coefficients show the long-run response of the dependent variable to the independent variables. The short-run coefficients include the error correction mechanism of the model. The error correction term measures how short-run imbalances are corrected in the long-run. If this term is negative and significant, the long-run equilibrium relationship is valid. The general form of the ARDL model is as in Equation 5:

$$Y_t = \alpha + \sum_{i=1}^{p} \beta_i Y_{t-i} + \sum_{i=0}^{q} \gamma_i X_{t-i} + \varepsilon_t \qquad (5)$$

In Equation 5, $Y_t$ is the dependent variable, $X_t$ is the independent variable, $p$ and $q$ are lag lengths, $\alpha$ is the constant term, $\beta_i$ and $\gamma_i$ are coefficients, and $\varepsilon_t$ is the error term. The ARDL method is advantageous as it provides reliable results even in small samples. It can also be used when the variables are I(0) or I(I). Moreover, the optimal lag lengths are determined using the Akaike Information Criterion (AIC) and Schwarz Information Criterion (SIC) proposed by Pesaran et al. (2001).

The ARDL method offers a significant advantage in that it allows for the use of integrated variables at both the I(0) and I(1) levels within the same model. This feature makes ARDL more flexible and practical compared to traditional cointegration tests, which require all variables in the dataset to be stationary at the same level. Additionally, its ability to produce reliable results even with small samples enhances the model's predictive power. The ARDL method allows for both dynamic and structural analysis by enabling the distinction between short-term and long-run relationships within the same equation. Once a cointegration relationship is identified, the error correction model (ECM) can be used to determine how deviations are corrected in the long-run.

**3.5. Robustness Check**

The coefficients obtained from the ARDL method are tested with FMOLS (Fully Modified Ordinary Least Squares) and CCR (Canonical Cointegrating Regression) methods. FMOLS and its methods are advanced techniques used to estimate the long-run cointegration relationship. The traditional ordinary least squares (OLS) estimator is susceptible to autocorrelation and variance problems. This can lead to biased and inconsistent estimates. To overcome these problems, the FMOLS method developed by Phillips and Hansen (1990) corrects the error terms for long-run properties. FMOLS produces degree-of-freedom corrected estimates considering autocorrelation and variance and provides reliable significance tests. Similarly, the CCR method proposed by Park (1992) allows the model to be rearranged to consider the correlation structure of the error terms. This method minimizes the impact of autocorrelation by transforming the long-run components of the variables. Thus, it allows for a more robust estimation of the coefficients. FMOLS and CCR methods have the advantage of providing efficient estimation especially in small samples. However, in order to apply these methods, the existence of cointegration relationship between variables should be tested beforehand.

Both methods have the advantage of producing more efficient, consistent, and unbiased estimates than classical methods, especially in small samples. FMOLS and CCR correct the estimation process by considering assumptions violations such as serial correlation and heteroscedasticity in the error terms, thereby increasing the reliability of parameter significance tests. Additionally, they strengthen the validity of empirical findings by enabling more robust statistical estimation of long-run relationships (Ugurlu & Keser, 2020; Ogul, 2022). In this regard, FMOLS and CCR methods play a complementary methodological role in testing the robustness of results obtained using the ARDL model.

## 4. EMPIRICAL FINDINGS

This section presents the empirical findings obtained to analyze the validity of jobless growth in Türkiye. In the first stage of the study, the stationarity levels of the variables used were determined. For this purpose, ADF and PP unit root tests are applied and the results are presented in Table 2.

**Table 2:** Unit Root Test Results

| Variable | | ADF unit root test | | | | PP unit root test | | | | Status |
|---|---|---|---|---|---|---|---|---|---|---|
| | | tau-statistic (level) | Lag | tau-statistic (first difference) | Lag | tau-statistic (level) | Lag | tau-statistic (first difference) | Lag | |
| UNP | Constant | -1.499 | [1] | -3.512** | [1] | -1.900 | [1] | -3.666*** | [2] | I(1) |
| AGR | | -1.613 | [0] | -5.422*** | [0] | -1.511 | [1] | -5.578*** | [1] | I(1) |
| IND | | -1.167 | [2] | -5.465*** | [1] | -1.120 | [2] | -3.485*** | [1] | I(1) |
| CON | | -2.037 | [1] | -3.108** | [0] | -1.242 | [1] | -3.147** | [0] | I(1) |
| SER | | -2.420 | [0] | -4.633*** | [0] | -2.420 | [0] | -4.628*** | [1] | I(1) |
| INF | | -1.276 | [0] | -4.242*** | [1] | -1.436 | [1] | -4.231*** | [2] | I(1) |
| UNP | Constant and Trend | -1.561 | [1] | -3.517** | [1] | -2.402 | [1] | -3.510** | [2] | I(1) |
| AGR | | -3.151 | [0] | -5.269*** | [0] | -3.185 | [1] | -5.405*** | [1] | I(1) |
| IND | | -1.666 | [2] | -5.552*** | [1] | -2.450 | [2] | -8.396*** | [1] | I(1) |
| CON | | -0.390 | [0] | -3.654** | [1] | -0.390 | [2] | -3.922** | [2] | I(1) |
| SER | | -2.271 | [0] | -4.580*** | [0] | -2.271 | [0] | -4.577*** | [1] | I(1) |
| INF | | -1.118 | [0] | -4.677*** | [1] | -1.201 | [1] | -3.316* | [2] | I(1) |

**Note: (1) The superscripts \*\*\*, \*\*, and \* denote the significance at a 1%, 5%, and 10% level, respectively. (2) The lag lengths were automatically selected by EViews.**

Table 2 presents the ADF and PP unit root test results. The ADF and PP test results show that the series are not stationary at level. However, all variables become stationary at first difference. Similar results are also observed in the tests including both constant and trend components. The first differences of the variables are stationary in both constant and, constant and trend models. Thus, it is concluded that all variables are I(I). This situation enabled the use of the ARDL method in the study. Therefore, the results obtained with the ARDL method are presented as follows.

**Table 3:** The Results of ARDL Bound Test

| Model | Optimal lag length | F-statistics | Critical values %5 | | Critical values %1 | |
|---|---|---|---|---|---|---|
| | | | $I(0)$ | $I(I)$ | $I(0)$ | $I(I)$ |
| F(UNP \| AGR, IND, CON, SER, INF) | (2, 2, 1, 1, 2, 2) | 5.557*** | 2.39 | 3.38 | 3.06 | 4.15 |

**Note: The superscripts ***, **, and * denote the significance at a 1%, 5%, and 10% level, respectively.**

Table 3 presents the results of the ARDL bound test. The results show that the F-statistic value is 5.557, which is above the upper critical thresholds at both 5% and 1% significance levels. This result indicates that there is a long-run cointegration relationship between the dependent variable UNP and the independent variables in the model. Simply stated, the variables move together in the long-run. Moreover, the optimal lag lengths are determined automatically using the information criteria (AIC, SC and HQ). This ensured that the model has the most appropriate lag structure. Thus, it allowed the dynamic relationships between variables to be modeled accurately. As a result, the results of the ARDL bounds test confirm the existence of a cointegration relationship between the variables in the model. Therefore, coefficient estimation is performed and the results are presented as follows.

**Table 4:** Short-run and Long-run Results

| Dependent variable: UNP$_{(M-D)}$ | Short-run coefficients | | |
|---|---|---|---|
| Regressors | **Coefficient** | **Std. Error** | **t-Statistic** |
| **AGR** | -0.471*** | 0.081 | -5.825 |
| **IND** | -0.680*** | 0.073 | -9.348 |
| **CON** | -0.899*** | 0.061 | -14.621 |
| **SER** | -1.383** | 0.206 | -2.338 |
| **INF** | -0.062*** | 0.014 | -4.369 |
| **ECT(-1)** | -0.118*** | 0.116 | -10.179 |
| Dependent variable: SDI$_{(M-D)}$ | Long-run coefficients | | |
| Regressors | **Coefficient** | **Std. Error** | **t-Statistic** |
| **AGR** | -2.380*** | 0.348 | -6.839 |
| **IND** | -4.057*** | 0.759 | -5.345 |
| **CON** | -1.761*** | 0.252 | -6.980 |
| **SER** | -3.664** | 1.000 | -2.377 |
| **INF** | -0.548** | 0.120 | -2.463 |
| C | 37.253*** | 7.213 | 5.165 |
| Diagnostic tests | | | *P value* |
| $\chi^2$ *(Serial correlation)* | | | 0.28 |
| $\chi^2$ *(Heteroskedasticity)* | | | 0.17 |
| $\chi^2$ *(Normality)* | | | 0.36 |
| $\chi^2$ *(Functional form)* | | | 0.47 |
| *CUSUM* | | | Stable |
| *CUSUMSQ* | | | Stable |

Note: The superscripts ***, **, and * denote the significance at a 1%, 5%, and 10% level, respectively.

Table 4 presents the ARDL estimation results. Table 4 shows that the error correction term (ECT) is negative and statistically significant, confirming the existence of a long-run equilibrium relationship among the variables in the model. The ECT coefficient value of -0.118 implies that approximately 12% of the short-run deviations from the long-run path are corrected in each period, indicating a relatively slow adjustment speed toward equilibrium.

The short-run coefficients show that all sectors have a negative and statistically significant impact on the unemployment rate. A one unit increase in the share of the agricultural sector in GDP decreases the unemployment rate by 0.471 points, 0.680 points in the industrial sector, 0.899 points in the construction sector and 1.383 points in the services sector in the short-run. The largest effect comes from the services sector, while the lowest effect belongs to the agricultural sector. These results reveal that the employment creation capacity in Türkiye differs across sectors. Industry and services sectors seem to have a larger impact on the labor market. The construction sector has long been recognized as one of the engines of economic growth in Türkiye. However, its employment creation potential has been found to be limited compared to other sectors. However, the agricultural sector traditionally has a lower employment capacity and tends to decline in labor demand with modernization and mechanization.

The long-run coefficients show that the impacts of sectors on unemployment are stronger in the long-run than in the short-run. A one-unit increase in the share of agriculture in GDP decreases the unemployment rate by 2.380 points, industry by 4.057 points, construction by 1.761 points and services by 3.664 points in the long-run. Industry and services sectors are critical for the labor market in the long-run. A growth policy based on the industrial sector in Türkiye can contribute to a further decline in the unemployment rate in the long-run. In particular, the strong negative impact of the industrial sector on unemployment shows that industrial policies are very important for employment creation. In developing countries like Türkiye, the development of industry both directly increases employment and supports the labor market through its indirect impact on other sectors. The services sector also plays an important role in reducing unemployment. In Türkiye, service sectors such as tourism, retail and finance constitute a major source of employment. However, it should be noted that jobs in the service sector are often low-paid and temporary.

Inflation (INF) has a negative impact on unemployment in the short-run. A one-unit increase in inflation decreases the unemployment rate by 0.062 points. This may support the Phillips curve, which explains the inverse relationship between inflation and unemployment in the short-run. In economies like Türkiye, where inflationary pressures are frequent, price increases may initially increase labor demand. However, whether this impact is sustainable in the long-run should be analyzed separately. In this study, the impact of inflation on unemployment in the long-run has become more evident. A one-unit increase in inflation reduces unemployment by 0.548 percentage points in the long-run. This result shows a similar relationship with the short-run. However, high inflation may create economic instability in the long-run and this relationship may change under different economic conditions. Employment-generating growth policies are important for reducing unemployment in Türkiye. Increasing economic diversification and supporting high value-added sectors also play a critical role.

The findings show that jobless growth is not generally valid in Türkiye. Economic growth reduces the unemployment rate. However, the capacity to create employment differs across sectors. Industry, construction and service sectors play an important role in this process. These sectors have a significant and strong impact in terms of employment creation potential. Especially industry and services sectors play a decisive role in the labor market in the long run. However, agriculture and construction sectors have a more limited impact on unemployment. In this regard, industry and services sectors should be prioritized in employment policies. High value-added sectors should be prioritized for sustainable employment. The labor market should adapt to this transformation.

The diagnostic tests applied to assess the validity of the model confirm that the model has no statistical problems. The results of the serial correlation test (p=0.28) show that there is no autocorrelation in the model and the forecasts are reliable. The variance test (p=0.17) reveals that the error terms have constant variance and the model exhibits a stable structure. Normality test (p=0.36) confirms that the error terms meet the assumption of normal distribution. The functional form test (p=0.47) shows that the model is correctly constructed and does not contain any functional deficiencies. Finally, the CUSUM and CUSUM Squared tests prove that the model estimates are stable over time and the estimation results are robust. The rest of the paper presents the robustness check results.

**Table 5:** Robustness Check Results

| Dependent variable: UNP | FMOLS | | | CCR | | |
|---|---|---|---|---|---|---|
| Variables | Coefficient | Std. Error | t-Statistic | Coefficient | Std. Error | t-Statistic |
| AGR | -1.470*** | 0.285 | -5.151 | -1.583*** | 0.296 | -5.349 |
| IND | -1.904*** | 0.593 | -3.211 | -2.082*** | 0.635 | -3.279 |
| CON | -1.229*** | 0.201 | -6.108 | -1.300*** | 0.197 | -6.546 |
| SER | -1.830* | 0.958 | -1.919 | -2.160** | 1.025 | -2.107 |
| INF | -0.104*** | 0.023 | -4.560 | -0.097*** | 0.028 | -3.418 |
| C | 20.886*** | 6.309 | 3.310 | 23.079*** | 6.424 | 3.593 |

Note: The superscripts ***, **, and * denote the significance at a 1%, 5%, and 10% level, respectively.

FMOLS and CCR methods are used to test the robustness of the findings obtained with the ARDL method. Table 5 presents the robustness test results. There are important similarities between ARDL, FMOLS and CCR estimations. First, the impact of all sectors on unemployment is negative and significant. This confirms that growth in agriculture, industry, construction, and services sectors reduces the unemployment rate as in the ARDL model. In all methods, the largest impact is observed in the industrial (IND) and services (SER) sectors. Moreover, the negative impact of inflation (INF) on unemployment is also consistent in both methods. These findings suggest that the relationship between unemployment and economic growth is consistent. The main results have not changed across different econometric methods.

Although the general trends are similar, there are some differences between the ARDL method and FMOLS and CCR. One of the most important differences is the estimated coefficient magnitudes. The coefficients of agriculture (AGR), industry (IND), construction (CON) and services (SER) sectors are estimated lower in FMOLS and CCR methods compared to the ARDL model. However, as a general trend, it has been consistently confirmed that the services and industry sectors have a strong mitigating impact on unemployment. This can be explained by the fact that the ARDL model considers both short-run and long-run impacts, while the FMOLS and CCR methods estimate only the long-run coefficients. The ARDL method is more able to address these fluctuations.

The findings of this study are consistent with the literature (e.g. Barisik et al., 2010; Ozdemir & Yildirim, 2013; Altuntepe & Guner, 2013), which emphasizes the reducing impact of growth on unemployment. However, this study provides a more detailed analysis by considering the impacts of sectoral growth on unemployment rates separately in the short and long-run. At this point, studies such as Ceylan and Sahin (2010), and Ozdemir and Yildirim (2013) have shown that the impact of growth on unemployment is more pronounced during periods of economic expansion and weakens during periods of contraction. However, in this study, the impacts of sectoral growth on unemployment are stronger in the long-run than in the short-run. The study is also consistent with studies emphasizing the importance of sectoral differences (e.g. Tuncer & Altiok, 2012; Aksoy, 2013; Abdioglu & Albayrak, 2015). It is also consistent with studies (e.g. Hanusch, 2013; Hamia, 2016; Jablanovic, 2017; Klinger and Weber, 2020) that indicate that sectoral structures and economic transformations are important factors shaping the relationship between growth and employment. In particular, the findings of the study emphasize the services and industrial sectors, contrary to studies that show that the employment generation capacity of the construction sector is more prominent (e.g. Abdioglu & Albayrak, 2015). In this respect, the study is consistent with studies showing that the increase in employment in the manufacturing sector remains limited (Tuncer & Altiok, 2012) and that the industrial and services sectors stand out (Aksoy, 2013).

## 5. CONCLUSION

This study examines the validity of the jobless growth hypothesis in Türkiye. For this purpose, the impact of sectoral growth on unemployment rate is analyzed using annual data for the period 2000-2022. ARDL method is used for cointegration and coefficient estimation. FMOLS and CCR methods are used to test the robustness of the findings obtained with the ARDL method.

The results of the analysis show that economic growth in Türkiye generally reduces the unemployment rate. In the period of the study, it is determined that the jobless growth hypothesis is not valid in Türkiye. In the short-run, all sectors have a negative and statistically significant impact on the unemployment rate. A one-unit increase in the share of the agricultural sector in GDP decreases the unemployment rate by 0.471 points, 0.680 points in the industrial sector, 0.899 points in the construction sector and 1.383 points in the services sector in the short-run. In the short-run, the largest impact comes from the services sector, while the lowest impact belongs to the agricultural sector. In the long-run, the impacts of sectors on the unemployment rate become stronger. A one unit increase in the share of the agricultural sector in GDP decreases the unemployment rate by 2.380 points, 4.057 points in the industrial sector, 1.761 points in the construction sector and 3.664 points in the services sector in the long-run. In the long-run, the industrial sector has the largest impact, while the construction sector has the smallest impact. These findings suggest that the industrial and service sectors play a determining role in the labor market. In particular, growth in the industrial sector has the highest capacity to reduce the unemployment rate in the long-run. It is understood that industry-based growth policies have a critical importance in creating employment in Türkiye. A shift to high value-added production will further reduce the unemployment rate. The services sector also offers ample employment opportunities. In particular, sub-sectors such as tourism, finance and retail play an important role in the labor market. Moreover, the inflation variable has a negative impact on unemployment in the short-run. A one-unit increase in inflation decreases the unemployment rate by 0.062 points. In the long-run, the impact of inflation on the unemployment rate becomes more pronounced. A one-unit increase in inflation decreases unemployment by 0.548 points. However, the possible risks of inflation on economic stability in the long-run should also be considered.

The findings of the study show that economic growth reduces the unemployment rate in Türkiye and the jobless growth hypothesis is not valid. Sectoral analysis reveals that especially the industrial and service sectors have a stronger impact on the unemployment rate. However, agriculture and construction sectors have a limited capacity to create employment. Although inflation seems to reduce the unemployment rate in the short-run, it should be kept in mind that it may pose a risk to economic stability in the long-run. In this perspective, the following policy recommendations are presented for Türkiye to develop job-creating growth policies and support sustainable employment: (1) The industrial sector has been found to be one of the sectors that reduce the unemployment rate the most in the long-run. Therefore, Türkiye should redirect its industrial policies towards high value-added production, increase its production capacity by expanding Industry 4.0 practices, and train qualified labor force. (2) The services sector is one of the largest sectors in Türkiye and offers a wide range of employment opportunities. Incentive mechanisms should be created to create more employment, especially in sub-sectors such as tourism, finance, IT and health. Entrepreneurship should be supported and vocational training programs should be increased. (3) The agricultural sector has a lower impact on reducing unemployment compared to other sectors. However, by increasing modernization and use of technology in agricultural production, productivity can be increased and employment can be created in rural areas. The economic contribution of the agricultural sector should be increased by supporting the integration of agriculture-food-industry. (4) The construction sector has long been recognized as one of the engines of economic growth in Türkiye. However, the employment generation capacity of the sector is generally limited to short-term projects. Therefore, investments should be made in infrastructure projects with long-term planning to create more sustainable employment in the construction sector. Training programs should be developed to ensure the transition of the labor force working in the sector to different sectors. (5) The study finds that inflation reduces the unemployment rate. However, high inflation may create economic instability in the long-term. Therefore, employment policies should be implemented in a balanced manner considering inflationary pressures. Price stability and employment targets should be harmonized. (6) Vocational and technical training programs should be strengthened to meet labor demand in industry and services sectors. Implement skills development training programs, especially for jobs based on digital economy, green energy and innovative technologies. (7) Develop sectoral policies and long-term strategies to ensure sustainable job creation. Priority should be given to transition to high value-added production, education reforms and job-creating growth policies to ensure lasting improvements in the labor market.

This study provides important findings on the validity of jobless growth in Türkiye. However, it has some limitations. Future studies can address these limitations. (1) This study assessed sectoral impacts on the unemployment rate. However, unemployment may have been affected differently in subgroups such as female employment, youth unemployment or regional unemployment. Future studies may consider these distinctions and conduct more detailed analyses. (2) The analysis is based on annual data for the period 2000-2022. Expanding the data set or using higher frequency data may contribute to a more detailed analysis of short and long-run impacts. (3) The model focuses on the main relationship between sectoral growth and unemployment. Other explanatory variables can be included in the model by expanding the data set. These limitations provide new opportunities for future research.